\documentclass[epj]{svjour}
\usepackage{epsfig}
\usepackage{times}
\usepackage{amsfonts}
\usepackage{graphics}
\usepackage{psfrag}
\usepackage{epsf}
\usepackage{float}
\usepackage{amssymb,stmaryrd,latexsym}
\usepackage{amsmath}
\usepackage{hhline}

\newcommand{\be}{\begin{equation}}
\newcommand{\ee}{\end{equation}}
\newcommand{\ba}{\begin{eqnarray}}
\newcommand{\ea}{\end{eqnarray}}
\newcommand{\non}{\nonumber}

\newcommand{\rang}{\right\rangle}

\begin{document}

\title{\hfill {\tiny FZJ-IKP-TH-2009-01, HISKP-TH-09/01}\\
Interactions between heavy mesons and Goldstone bosons from chiral
dynamics}
\author{
Feng-Kun Guo\inst{1}\thanks{{\it Email address:}
f.k.guo@fz-juelich.de}, Christoph Hanhart\inst{1,2}\thanks{{\it
Email address:} c.hanhart@fz-juelich.de}, and Ulf-G.
Mei{\ss}ner\inst{1,2,3}\thanks{{\it Email address:}
meissner@itkp.uni-bonn.de}} \institute{Institut f\"{u}r Kernphysik
and J\"ulich and Center for Hadron Physics, Forschungszentrum
J\"{u}lich, D--52425 J\"{u}lich, Germany \and Institute for Advanced
Simulations, Forschungszentrum J\"ulich, D--52425 J\"{u}lich,
Germany \and Helmholtz-Institut f\"ur Strahlen- und Kernphysik and
Bethe Center for Theoretical Physics, Universit\"at Bonn, D--53115
Bonn, Germany}
\date{Received: date / Revised version: date}


\abstract{
We calculate the S-wave scattering lengths for charmed mesons
scattering off Goldstone bosons and explore their quark mass
dependence using chiral perturbation theory up to next-to-leading
order as well as a unitarized version of it. The quark mass
dependence of all scattering lengths determined in a recent lattice
calculation can be reproduced by the unitarized version. We also
discuss signals of possible bound states in these observables. }


\authorrunning{F.-K.~Guo {\it et al.}}
\titlerunning{Interactions between heavy mesons and Goldstone bosons}

%

\maketitle

\section{Introduction}

The interaction strengths amongst hadrons are fundamental quantities
of the strong interactions. There are two different approaches to
analyze the low--energy scattering amplitudes in a
model--independent manner. First, one can use the  chiral
perturbation theory (ChPT) representation of the pertinent
scattering amplitude which is given in terms of a few LECs. These
might either be determined directly from experiment or from full QCD
calculations performed on a discretized space--time lattice. A
beautiful example in this context is the accurate  determination of
the pionic LECs $l_3$ and $l_4$ from lattice simulations --- see
Refs.~\cite{latticepaper} and references therein. The second
approach to address the issue of scattering lengths is the
exploitation of L\"uscher's formula
\cite{Luscher:1986pf,Luscher:1990ux} in the analysis of the lattice
gauge theory results. It allows one to relate two--particle
scattering lengths to the behaviour of certain energy levels in a
finite volume. In full lattice QCD, the maximally stretched states
for $\pi \pi$, $\pi K$ and $KK$ scattering have been systematically
and accurately analyzed by the NPLQCD collaboration based on an
extension of the L\"uscher formula (for a review, see
Ref.~\cite{Beane:2008dv} and references therein).

In this paper we will discuss another sector, where the same program
can be carried out, namely the scattering of Goldstone bosons off
$D$--mesons. A recent lattice study for these systems is given in
Ref.~\cite{Liu:2008rz}. However, contrary to the $\pi\pi$ system,
here an additional complication may occur, for the interaction in
some channels might be sufficiently strong to show non-perturbative
phenomena even on the level of hadron--hadron scattering, eventually
leading to hadron bound states --- the so--called hadronic
molecules. A famous example of relevance here is the
$D_{s0}^*(2317)$ identified as a $DK$ bound state in various
studies~\cite{eef,Kolomeitsev:2003ac,Guo:2006fu,oset}. We will
discuss to what extent non-perturbative phenomena might already be
present in the channels investigated on the lattice and also point
out the most interesting channels not yet studied. Note also that a
detailed knowledge of Goldstone boson--$D$-meson scattering is not
only of interest on its own, it is also an essential input for
calculations of $D$--mesons in matter~\cite{oset2}.

Preceding the direct lattice calculations of the
 Goldstone boson--$D$-meson scattering lengths,
the one for the  $I=1/2$ $D\pi$ channel (where $I$ denotes the total
isospin)  was extracted using the lattice
calculations of the $D\pi$ scalar form factor in semileptonic
$D\to\pi$ decays in Ref.~\cite{Flynn:2007ki}. The result was $M_\pi
a= 0.29(4)$, i.e. $a=0.41(6)$~fm (where $a$ denotes the scattering length).
Using the same method, the value
of the isoscalar $DK$ scattering length was found to be effectively
infinite, which was interpreted as a signal of a bound state of $DK$ at threshold.

In this paper, we present a systematic study of the S-wave
scattering lengths of the Goldstone boson--$D$-meson interactions
for all possible channels. In Section~\ref{sec:slphy}, the results
in the physical world (i.e. at the physical values of the quark
masses) are calculated to leading order (LO) and next-to-leading
order (NLO) in the chiral expansion, as well as those obtained using
unitarized scattering amplitudes. In Section~\ref{sec:chextra}, we
extrapolate the results to unphysical quark masses using the
unitarized scattering amplitudes. Some discussions and a brief
summary is given in Section~\ref{sec:sum}. Some useful isospin
relations are relegated to Appendix~\ref{app:isore}.

\vfill

\section{Scattering lengths at the physical quark masses}
\label{sec:slphy}
\subsection{The chiral effective Lagrangians} \label{sec:lag}

The leading order Lagrangian is just given by the kinetic and the mass
term of the heavy mesons (chirally coupled to
pions)~\cite{Burdman:1992gh,Wise:1992hn,Yan:1992gz}
\begin{equation}
{\cal L}^{(1)} = {\cal D}_{\mu}D{\cal
D}^{\mu}D^{\dag}-\overset{_\circ}{M}_D^2 DD^{\dag}
\end{equation}
with $D=(D^0,D^+,D_s^+)$ collecting the charmed mesons with the chiral
limit mass $\overset{_\circ}{M}_D$. The covariant derivative is
\begin{eqnarray}
{\cal D}_{\mu} &\!\!=&\!\! \partial_{\mu}+\Gamma_{\mu}, \nonumber\\
\Gamma_{\mu} &\!\!=&\!\!
{\frac12}\left(u^{\dag}\partial_{\mu}u+u\partial_{\mu}u^{\dag}\right),
\end{eqnarray}
where
\ba%
U = \exp \left( \frac{\sqrt{2}i\phi}{F}\right),\quad u^2=U~,
\ea%
with $F$ the Goldstone boson decay constant in the chiral limit (which we
will identify with the pion decay constant in
the following). The Goldstone bosons are collected in the matrix-valued field
\ba%
\label{eq:phi}
 \phi =
  \left(
    \begin{array}{c c c}
 \frac{1}{\sqrt{2}}\pi^0+\frac{1}{\sqrt{6}} \eta & \pi^+ & K^+ \\
\pi^- & - \frac{1}{\sqrt{2}}\pi^0+\frac{1}{\sqrt{6}} \eta & K^0 \\
K^- & \bar K ^0 & -\frac{2}{\sqrt{6}} \eta \\
    \end{array}
\right) .
\ea%

Since we are only interested in the scattering lengths with definite
isospin, we only consider the strong interaction part of the NLO
chiral Lagrangian describing the interactions of the pseudoscalar
charmed mesons with the Goldstone bosons, which
reads~\cite{Guo:2008gp}
\ba%
\label{eq:L2str} {\cal L}^{(2)}_{\rm str.} &\!\!=&\!\! D \bigl(
-h_0\langle\chi_+\rangle - h_1\tilde{\chi}_+ + h_2\left\langle
u_{\mu}u^{\mu} \right\rangle - h_3u_{\mu}u^{\mu}
\bigr) {\bar D} \non\\
& + & {\cal D}_{\mu}D \bigl( h_4\langle u^{\mu}u^{\nu}\rangle - h_5
\{u^{\mu},u^{\nu}\} - h_6 [u^{\mu},u^{\nu}] \bigr) {\cal
D}_{\nu}{\bar D} , \nonumber\\ &&
\ea%
 where
\ba%
\chi_+ &\!\!=&\!\! u^\dagger \chi u^\dagger + u\chi u ,\nonumber\\
\tilde{\chi}_+ &\!\!=&\!\! \chi_+ -
{\frac13}\left\langle\chi_+\right\rangle,\nonumber\\
u_{\mu} &\!\!=&\!\! iu^{\dag}{\cal D}_{\mu}Uu^{\dag}.
\ea%
The quark mass matrix is diagonal
\begin{equation}
\chi = 2B\cdot {\rm diag}\left(m_u,m_d,m_s\right)~,
\end{equation}
in terms of $B= |\langle 0 |\bar q q |0\rangle|/F^2 $.
The unknown coefficients $h_i (i=0, \ldots ,6)$ are the pertinent
LECs. At ${\mathcal O}(p^2)$, $h_1$ can be determined
from the mass differences among the $D$--mesons as
$h_1=0.42$~\cite{Guo:2008gp}. In the following, we choose to drop
the terms with one more flavor trace in the NLO Lagrangian, i.e. the
$h_0$, $h_2$ and $h_4$ terms, which are suppressed in the large
$N_C$--limit of QCD~\cite{Manohar:1998xv,Lutz:2007sk}. The $h_6$ term
is also dropped since it is suppressed by one order due to the
commutator structure. We are therefore left with only two free,
active parameters, namely $h_3$ and $h_5$. It can be shown that the
contributions of the $h_5$ and $h_3$ terms to S-wave amplitudes
differ only at ${\mathcal O}(p/m_D)$.
Therefore a variation of $h_3$ within its natural
bounds for a given $h_5$ provides an estimate for
higher order contributions~\cite{Guo:2008gp}.

\subsection{Numerical results for the scattering lengths}
\label{sec:res}

\begin{table*}[t]
\begin{center}
\renewcommand{\arraystretch}{1.2}
\begin{tabular}{|ll|lll|}\hline\hline
$(S,I)$  & Channel & $C_0$ & $C_1$ & $C_{35}$  \\ \hline%
$(-1,0)$ & $D{\bar K}\to D{\bar K}$ & $-1$ & $5M_K^2$ & $-1$ \\%
$(-1,1)$ & $D{\bar K}\to D{\bar K}$ & $\;\;\:1$ & $-M_K^2$ & $\;\;\:1$ \\%
$\left(0,{\frac12}\right)$ & $D\pi\to D\pi$ & $-2$ & $-M_\pi^2$ & $\;\;\:1$ \\%
                           & $D\eta\to D\eta$ & $\;\;\:0$ & $2M_\eta^2-M_\pi^2$ & $\;\;\:{\frac13}$ \\%
                           & $D_s{\bar K}\to D_s{\bar K}$ & $-1$ & $-M_K^2$ & $\;\;\:1$ \\%
                           & $D\eta\to D\pi$ & $\;\;\:0$ & $-3M_\pi^2$ & $\;\;\:1$ \\
                           & $D_s{\bar K}\to D\pi$ & $-\frac{\sqrt{6}}{2}$ & $-\frac{3\sqrt{6}}{4}\left(M_K^2+M_\pi^2\right)$ & $\;\;\:\frac{\sqrt{6}}{2}$ \\
                           & $D_s{\bar K}\to D\eta$ & $-\frac{\sqrt{6}}{2}$ & $\frac{\sqrt{6}}{4}\left(5M_K^2-3M_\pi^2\right)$ & $-\frac{\sqrt{6}}{6}$ \\
$\left(0,{\frac32}\right)$ & $D\pi\to D\pi$ & $\;\;\:1$ & $-M_\pi^2$ & $\;\;\:1$ \\%
$(1,0)$ & $DK\to DK$ & $-2$ & $-4M_K^2$ & $\;\;\:2$ \\%
        & $D_s\eta\to D_s\eta$ & $\;\;\:0$ & $-2\left(2M_\eta^2-M_\pi^2\right)$ & $\;\;\:{\frac43}$ \\%
        & $D_s\eta\to DK$ & $-\sqrt{3}$ & $-\frac{\sqrt{3}}{2}\left(5M_K^2-3M_\pi^2\right)$ & $\;\;\:\frac{\sqrt{3}}{3}$ \\
$(1,1)$ & $D_s\pi\to D_s\pi$ & $\;\;\:0$ & $2M_\pi^2$ & $\;\;\:0$ \\%
        & $DK\to DK$ & $\;\;\:0$ & $2M_K^2$ & $\;\;\:0$ \\%
        & $DK\to D_s\pi$ & $\;\;\:1$ & $-{\frac32}\left(M_K^2+M_\pi^2\right)$ & $\;\;\:1$ \\
$(2,\frac12)$ & $D_sK\to D_sK$ & $\;\;\:1$ & $-M_K^2$ & $\;\;\:1$ \\\hline\hline%
\end{tabular}
\caption{\label{tab:Vstu}The coefficients in the scattering
amplitudes. Here, $S$ ($I$) denotes the total strangeness (isospin)
of the two--meson system.}
\end{center}
\end{table*}

We first consider the perturbative chiral expansion of the scattering
amplitude, $T = T^{(1)} + T^{(2)} + \ldots$, where the superscript
denotes the chiral dimension. We work here to NLO, that is chiral
dimension two. In that case, the scattering amplitudes are real and
can be written as
\ba%
\label{eq:vstu} T (s,t,u) &=&  T^{(1)} (s,t,u) + T^{(2)} (s,t,u)\nonumber\\
&=& \frac{C_0}{4F^2}(s-u) + \frac{2C_1}{3F^2}h_1
+ \frac{2C_{35}}{F^2}H_{35}(s,t,u),
\ea%
with \be H_{35}(s,t,u)=h_3\, p_2\cdot p_4+h_5 \, (p_1\cdot
p_2p_3\cdot p_4+p_1\cdot p_4p_2\cdot p_3)~. \ee The coefficients in
all the amplitudes for the Goldstone boson--$D$-meson scattering are
given in Table~\ref{tab:Vstu}. The NLO S-wave elastic scattering
amplitudes at threshold of the participating particles can be cast
into the form
\ba%
T_{\rm thr} &=& \frac{1}{F^2}\biggl[C_0M_1M_2 + \frac{2C_1}{3}h_1 +
2C_{35}\bigl(h_3M_2^2\nonumber\\
&& \qquad\qquad + 2h_5M_1^2M_2^2\bigr) \biggr],
\ea%
with $M_1$ and $M_2$ denoting the masses of the scattered heavy and
light mesons, respectively. Note that from Table~\ref{tab:Vstu} one
can see that the most attractive interaction occurs in the
$(S,I)=(1,0)$ $DK$ channel, where $S$ ($I$) denotes the total
strangeness (isospin) of the two--meson system. This is the reason
why in this channel the $D_{s0}^*(2317)$ was generated dynamically
in many previous
works~\cite{eef,Kolomeitsev:2003ac,Guo:2006fu,oset}.

The S-wave scattering length parameterizes the scattering amplitude
at threshold
\be%
a_0 = -\frac1{8\pi(M_1+M_2)}T_{\rm thr} ~.
\ee%
Up to the order we are working, $F$ can be replaced by the physical
pion decay constant $F_\pi=92.4$~MeV. We take the physical masses
for all the mesons, i.e., $M_\pi=138$~MeV, $M_K=496$~MeV,
$M_\eta=548$~MeV, $M_D=1867$~MeV, and
$M_{D_s}=1968$~MeV~\cite{Amsler:2008zz}. The results for the S-wave
scattering lengths at LO are given in the third column of
Table~\ref{tab:a0}.

\begin{table*}[t]
\begin{center}
\renewcommand{\arraystretch}{1.3}
\begin{tabular}{|ll|llll|r|}\hline\hline
$(S,I)$  & Channel & LO & NLO & UChPT & CUChPT & Lattice~\cite{Liu:2008rz}~ \\ \hline%
$(-1,0)$ & $D{\bar K}\to D{\bar K}$ & $\;\;\:0.36$ & $\;\;\:0.31(2)$ & $\;\;\:0.96(20)$& &\\
$(-1,1)$ & $D{\bar K}\to D{\bar K}$ & $-0.36$ & $-0.41(2)$ & $-0.22(2)$ &   &
$-0.23(4)$ \\
$\left(0,{\frac12}\right)$ & $D\pi\to D\pi$ & $\;\;\:0.24$ & $\;\;\:0.23(0)$ &
$\;\;\:0.36(1)$ & $\;\;\:0.35(1)$ &  \\
                           & $D\eta\to D\eta$ & $\;\;\:0$ & $-0.09(1)$ &
                           $-0.08(1)$ & $\;\;\:0.19(9)+i0.02(2)$ & \\
                           & $D_s{\bar K}\to D_s{\bar K}$ & $\;\;\:0.36$ &
                           $\;\;\:0.31(6)$ & $\;\;\:1.10(57)$ &
                           $-0.60(53)+i0.77(15)$& \\
$\left(0,{\frac32}\right)$ & $D\pi\to D\pi$ & $-0.12$ & $-0.12(0)$ & $-0.10(1)$ &   & $-0.16(4)$ \\
$(1,0)$ & $DK\to DK$ & $\;\;\:0.72$ & $\;\;\:0.67(4)$ & $-1.47(20)$ &
$-0.93(5)$ & \\
        & $D_s\eta\to D_s\eta$ & $\;\;\:0$ & $\;\;\:0.00(10)$ & $\;\;\:0.02(10)$ & $-0.33(4)+i0.05(1)$ &\\
$(1,1)$ & $D_s\pi\to D_s\pi$ & $\;\;\:0$ & $-0.005$ & $-0.005$ & $-0.0003(4)$
& $\;\;\:0.00(1)$ \\
        & $DK\to DK$ & $\;\;\:0$ & $-0.054$ & $-0.049$ & $-0.04(6)+i0.29(11)$ & \\
$(2,{\frac12})$ & $D_sK\to D_sK$ & $-0.36$ & $-0.41(6)$ & $-0.23(5)$ &   &
$-0.31(2)$~
\\\hline\hline%
\end{tabular}
\caption{\label{tab:a0}The S-wave scattering lengths from
calculations at LO and NLO (units are fm). The results using
unitarized amplitudes are also given in the two columns denoted by
UChPT and CUChPT, representing one--channel and coupled--channel
unitarized chiral perturbation theory, respectively.}
\end{center}
\end{table*}

To work out the results to NLO, we take the dimensionless low--energy
constant $h_5'\equiv h_5 M_{D^0}^2$ to be in the natural range of
$[-1,1]$ as in Ref.~\cite{Guo:2008gp}.\footnote{There is a typo in
  Ref.~\cite{Guo:2008gp}. There $h_5'$ was written as $h_5/M_{D^0}^2$
  while the correct one should be $h_5 M_{D^0}^2$.}  Correspondingly,
$h_3$ is determined from fitting to the mass of the $D_{s0}^*(2317)$
in the full calculation, to be described below.  This leads to
$h_3=-1.479$ for $h_5'=1$, and $h_3=2.315$ for $h_5'=-1$. The
results for the S-wave scattering lengths using these input
parameters also for the perturbative calculation
 to NLO are given in the fourth
column of Table~\ref{tab:a0} with the uncertainties from the lack of
knowledge of $h_3$ and $h_5'$. We use these uncertainties as an
estimate for possible higher order corrections. A comparison of the
NLO results with the LO ones shows a good convergence of the chiral
expansion, especially for channels where only $D$-mesons and pions
are involved. Even in the channels with kaons or etas, no dramatic
NLO corrections are found.
\begin{table*}[t]
\begin{center}
\renewcommand{\arraystretch}{1.3}
\begin{tabular}{|lll|ccc|ccc|}\hline\hline
$(S,I)$  & Channel & Thr &  & $h_5'{=+}1$ & & &  $h_5'{=-}1$  &  \\
         &         &           & Re & Im & RS & Re & Im & RS \\
 \hline%
$(-1,0)$ & $D{\bar K}\to D{\bar K}$ & 2363 & 2354 & $\pm$56 & II & 2301 & $\pm$97 & II \\
$\left(0,{\frac12}\right)$ & $D\pi\to D\pi$ &
2005 & 2098 & $\pm$124 & II & 2102 & $\pm$106 & II \\
                           & $D_s{\bar K}\to D_s{\bar K}$ &
2464 & 2286 & $\pm$54 & II & 2354 & 0 & II \\
& &     &      &     &    & 2431 & 0 & II \\
$(1,0)$ & $DK\to DK$ & 2363 & 2343 & 0 & I & 2337 & 0 & I \\
\hline\hline%
\end{tabular}
\caption{\label{tab:poles1}Real parts, imaginary parts
and Riemann sheet (RS) of the pole positions for the one--channel
calculations for two parameter sets. All masses/energies are given in MeV.}
\end{center}
\end{table*}
\begin{table*}[t]
\begin{center}
\renewcommand{\arraystretch}{1.3}
\begin{tabular}{|l|ccc|ccc|}\hline\hline
$(S,I)$   &  & $h_5'{=+}1$ & & &  $h_5'{=-}1$  &  \\
                  & Re & Im & RS & Re & Im & RS \\
 \hline%
$\left(0,{\frac12}\right)$ & 2107 & $\pm$123 & II & 2107 & $\pm$105 & II \\
                           & 2452 & $\pm$17  & III & 2519 & $\pm$69 & III \\
$\left(0,{\frac12}\right) \ (V_{ii}=0)$ & 2466 & $\pm$24 & III & 2388 & $\pm$49 & III \\
$(1,0)$ & 2318 & 0 & I & 2318 & 0 & I \\
$(1,1)$ & 2309 & $\pm111$ & III & 2283 & $\pm196$ & III\\
\hline\hline%
\end{tabular}
\caption{\label{tab:poles2}Positions of poles with the largest
impact on physical observables from the coupled channel calculations
for two parameter sets. $V_{ii}=0$ denotes the results for a
calculation where the diagonal interactions were switched off. Here
the second (third) Riemann sheet for the $(0,1/2)$ case is defined
by Im($q_{D\pi}$)$<$0, Im($q_{D\eta}$)$>$0, and Im($q_{D_s\bar
K}$)$>$0 (Im($q_{D\pi}$)$<$0,  Im($q_{D\eta}$)$<$0, and
Im($q_{D_s\bar K}$)$>$0), and that for the $(1,1)$ case is defined
by Im($q_{D_s\pi}$)$<$0, and Im($q_{DK}$)$>$0 (Im($q_{D_s\pi}$)$<$0,
and Im($q_{DK}$)$<$0). All masses/energies are given in MeV.}
\end{center}
\end{table*}

The above calculations are performed with perturbation theory up to
a given order. However, a perturbative expansion to a finite order
can never be reliable, if there is a bound state, a virtual state,
or a resonance near by. All of these are non-perturbative phenomena,
and the presence of such kind of state would  modify the results
from perturbation theory significantly. Therefore, we unitarize the
amplitudes obtained from ChPT according to the method of
Ref.~\cite{Oller:2000fj} (called UChPT in the following. An early
review on this method is \cite{Oller:2000ma}). The unitarized
amplitude is then given by the following resummation
\begin{equation}
\label{eq:uni} T(s)=V(s)\left[1-G(s)\cdot V(s)\right]^{-1},
\end{equation}
with $V(s)$ the S-wave projection of the scattering amplitude $T =
T^{(1)} + T^{(2)}$ given in Eq.~(\ref{eq:vstu}). Further, $G(s)$ is
the scalar two--meson loop integral~\cite{Oller:2000fj,Oller:1998zr}
regularized by a subtraction constant $a(\mu)$ with $\mu$ denoting
the scale of the dimensional regularization.
The expression of
the loop integral at threshold is rather simple
\ba%
G(s_{\rm thr}) &=& \frac{1}{16\pi^2} \biggl[a(\mu) + \frac{1}{M_1+M_2} \nonumber\\
&& \quad \times\left(M_1\ln{\frac{M_1^2}{\mu^2}} +
M_2\ln{\frac{M_2^2}{\mu^2}}\right)\biggr]\, .
\ea%
We will use the subtraction constant determined in
Ref.~\cite{Guo:2008gp}, i.e. $a(1~{\rm GeV})=-1.846$. Then the
scattering lengths with the unitarized amplitude for each channel
can be obtained easily, and the results are given in the column
denoted by UChPT in Table~\ref{tab:a0}. For the channels with
repulsive interactions, i.e. those with negative scattering lengths
at LO and NLO in the convention used here, the correction from the
resummation is not dramatic. It is several percent for the channels
without kaons, and 50\% for the channels with kaons at most. For the
channels with attractive interactions, i.e. those with positive
scattering lengths at LO and NLO, the smallest correction is already
50\% for the $(0,1/2)$ $D\pi$ channel. Both the scattering lengths
of the $(-1,0)$ $D{\bar K}$ and  $(0,1/2)$ $D_s\bar K$ channels
change from about 0.3~fm to about 1~fm, and that of the $(1,0)$ $DK$
even changes the sign.
These dramatic changes can be understood on the
basis of the singularity structure of the unitarized
amplitudes.

\begin{figure*}[t]
\begin{center}
\epsfig{file=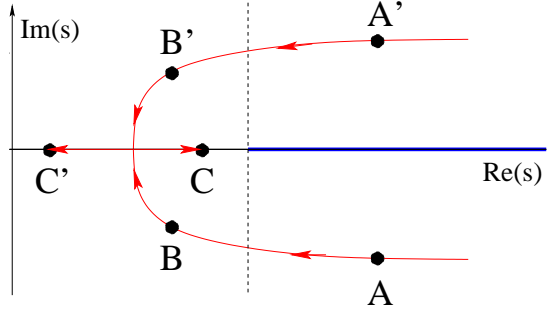,width=0.4\textwidth} \caption{Sketch
of typical trajectories of $S$--matrix poles in the second Riemann
sheet of the complex $s$--plane for energy-dependent potentials when
some strength parameter is changed. See text for meaning of the
labels. The dashed vertical line indicates the position of the
elastic scattering threshold and the thick horizontal line the
resulting unitarity cut. \label{fig:polemovement}}
\end{center}
\end{figure*}

At this point some general remarks are in order.  Imagine some
energy--independent potential that provides an attractive force
between two particles.  As the strength of the force is increased,
eventually a virtual state appears in the unitarized scattering
amplitude. A virtual state is a singularity below threshold on the
real axis on the second sheet of the complex $s$--plane of the
S--matrix. As the interaction strength is further increased the pole
moves closer and closer to the threshold. At the same time the
scattering length grows to become positive infinity when the pole
hits the threshold. This situation is almost realized in $nn$
scattering near threshold which shows an exceptionally large
scattering length of about $18$ fm. If the strength of the
interaction is increased even further, the pole jumps on the
physical sheet --- the state becomes a bound state. Correspondingly
the scattering length changes sign and starts to increase from
negative infinity as the pole moves away from the threshold --- this
is the case that applies to the deuteron channel of nucleon--nucleon
scattering, where the scattering length is about $-5$ fm (Note that
the sign conventions in nucleon--nucleon scattering are often
different to those used here). These scenarios are discussed in
various textbooks, see e.g. Refs.~\cite{taylor,McVoy}.  For
energy--dependent interactions one may even start from a resonance,
which corresponds to two poles in the second Riemann sheet, both
located as mirror images with respect to reflections on the real
axis. This is illustrated as the points A and A' in
Fig.~\ref{fig:polemovement}.  Then an increase of the interaction
strength will let the magnitude of both the real part and the
imaginary part decrease and will eventually lead to poles located
below the elastic threshold, but still with non-vanishing imaginary
parts (B and B' in the figure). A further increase in the strength
of the potential might convert the two resonance poles into two
virtual states (C and C' in the figure). As the strength parameter
gets increased even further, one pole moves towards the threshold to
eventually become a bound state while the other one moves further
away from the physical regime --- this behaviour was e.g. reported
in Ref.~\cite{extrapol} for the case of the light scalar meson
$f_0(600)$, where the interaction strength was varied by a variation
of the pion mass and was also observed in phenomenological
studies~\cite{eef2}.  We checked that indeed the mentioned types of
poles appear by the unitarization. As can be read off
Table~\ref{tab:poles1}, showing the pole positions for the single
channel calculations, for $h_5'=+1$ in the $D_s\bar K$ channel there
appears a pair of resonance poles below the $D_s\bar K$ threshold.
For the other parameter set displayed in the table ($h_5'=-1$) those
changed into two virtual states. For the full calculation with
coupled channels for both parameter sets there are only resonance
poles. In addition there are pairs of resonance poles in the second
Riemann sheet for the channels $(0,1/2)$ $D\pi$ and $(-1,0)$ $D\bar
K$. The former one was found already previously in
Ref.~\cite{Kolomeitsev:2003ac}. Note also that our findings for the
scattering length in that  channel are consistent with those of
Ref.~\cite{Flynn:2007ki} within the theoretical uncertainty.  In the
$(1,0)$ $DK$ even a bound state appears. As a consequence the
scattering length changes its sign.
In Refs.~\cite{Weinberg:1965zz} in was shown that for bound states
near a threshold one may write for the scattering length
\be%
a=-2\left(\frac{1-Z}{2-Z}\right)\frac{1}{\sqrt{2\mu\epsilon}}\left(1 + {\cal
  O}(\sqrt{2\mu\epsilon}{}/\beta)\right) \ .
\ee%
Applied to the $D_{s0}^*(2317)$ we have
$\epsilon=M_D+M_K-M_{D_{s0}^*(2317)}$ for the binding energy,
$\mu=M_D M_K/(M_D+M_K)$ for the reduced mass, and $1/\beta$, the
range of forces, may be estimated via $\beta \sim m_\rho$.  The
quantity $Z$, which can be identified with the wave function
renormalization constant, is a measure of the molecular component of
the state, with $Z=1$ ($Z=0$) for a pure elementary (molecular)
state.
 Taking $Z=0$, the above equation gives
$a^{(1,0)}_{DK\to DK}=-1.05$~fm, which is close to the value listed
in Table~\ref{tab:a0}. Thus, would this value be extracted from
lattice simulations in the future, it would be  a direct proof for
the molecular nature of the $D_{s0}^*(2317)$.


For $(S,I)=(0,1/2)$, $(1,0)$ and $(1,1)$, there are more channels
with the same quantum numbers, and the unitarization should include
also the coupled--channel effect (called CUChPT for simplicity). The
off--diagonal interactions can induce an imaginary part to the
scattering length for those channels coupling to a channel with a
lower threshold. The results with coupled--channel unitarization are
given in the column denoted by CUChPT in Table~\ref{tab:a0}. From
the amplitude coefficients for $(S,I)=(0,1/2)$ in
Table~\ref{tab:Vstu}, one finds the off--diagonal interactions for
the $D_s\bar K$ to $D\pi$ and $D\eta$ are large.  As can be seen
from the line labled by $V_{ii}=0$ in Table~\ref{tab:poles2} they
alone can already produce poles. For the full calculation their
effect is to move the poles on the second Riemann sheet of the
one--channel $D_s\bar K\to D_s\bar K$ amplitude to a pole in the
relative first Riemann sheet of the $D_s\bar K$ channel (that is the
Riemann sheet with positive imaginary part of the center-of-mass
system (cms) three--momentum in this channel and negative cms
three--momentum in the other two channels --- for the full
calculation this sheet is labled as III, for by convention the first
sheet is related to the channel with the lightest threshold). The
appearance of the pole on that physical sheet again leads to a
change in sign of the real parts of the scattering lengths for both
the $D\eta$ and $D_s\bar K$ channels. For $(S,I)=(1,1)$, the LO
diagonal interactions vanish, and the NLO ones are repulsive. The
off--diagonal interaction however produces a pole which is shown in
the last line in Table~\ref{tab:poles2}. Thus the pole is purely a
coupled--channel effect. Because of the pole the imaginary part of
the scattering length in the $(1,1)$ $DK$ channel becomes much
larger than its real part, a direct reflection of the
coupled--channel effect.

\begin{figure*}[t]
\begin{center}
\psfrag{ch1}{$(-1,0)~D{\bar K}\to D{\bar K}$}
\psfrag{ch2}{$(-1,1)~D{\bar K}\to D{\bar K}$}
\psfrag{ch3}{$(0,{\frac32})~D\pi\to D\pi$}
\psfrag{ch4}{$(2,{\frac12})~D_sK\to D_sK$}
\epsfig{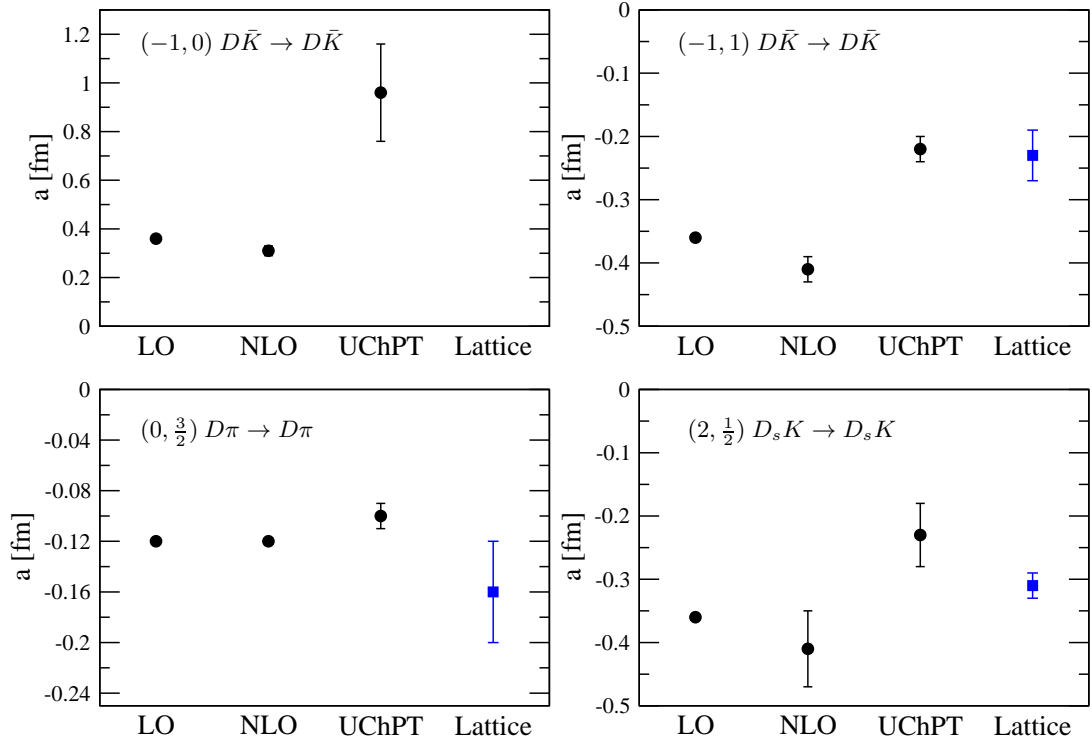}
\caption{Comparison of the scattering lengths
in four channels with the lattice data. We give the
results for LO, NLO and unitarized ChPT.
\label{fig:slphy}}
\end{center}
\end{figure*}
There are no experimental data for the scattering of Goldstone
bosons off $D$--mesons available. However, for four of the channels
discussed above results of lattice gauge theory calculations were
recently published~\cite{Liu:2008rz}.%
These lattice data have to be taken with a grain of salt as they
have been obtained for a single lattice spacing and volume, thus
leaving room for sizeable systematic discretization effects.%
The resulting scattering lengths are also shown in
Table~\ref{tab:a0}. To illustrate the level of agreement for the LO,
NLO, and UChPT calculations with the lattice data the results for
three of the channels are also shown in Fig.~\ref{fig:slphy} (so far
lattice data exist only for channels without channel coupling). For
the fourth channel investigated on the lattice the scattering length
is very small as it is for all the different calculations we
performed. Instead of this channel, in the figure we show the
channel $(-1,0)~D{\bar K}\to D{\bar K}$, which could be easily
investigated on the lattice with the same tools already employed in
Ref.~\cite{Liu:2008rz}.  The first and important observation is that
the NLO corrections are generally small --- the series appears to be
well behaved. In addition, the figure shows that for two of the
channels, {$(0,{\frac32})~D\pi\to D\pi$} and {$(2,{\frac12})~D_sK\to
  D_sK$}, the agreement of the calculations with the lattice results
is of similar quality for all three schemes, however, for the
channel $(-1,1)~D{\bar K}\to D{\bar K}$ an agreement with the
lattice results is achieved only after unitarization.  This is a
priori unexpected, since the interaction in this channel is
repulsive. However, a repulsive interaction iterated to even orders
gives an attractive interaction and consequently by the
unitarization the repulsion gets weakened --- the scattering length
gets smaller in magnitude. We checked that in our calculation the
leading loop, formally a next-to-next-to-leading order contribution,
already gives the bulk of the effect.  Thus here the lattice results
clearly show the trace of a higher order hadron--hadron interaction.
As the first panel of the figure shows, we predict an even more
dramatic effect in the channel $(-1,0)~D{\bar
  K}\to D{\bar K}$. As discussed above, here the interaction is
attractive and the large change in the scattering length is the
consequence of the appearance of a resonance state in the scattering
matrix --- clearly a non-perturbative phenomenon.  It would
therefore be of high theoretical interest to have lattice results
for this channel as well.

Ideally lattice simulations would be available for all channels
calculated. As can be read off Table~\ref{tab:a0}, in some channels
the effect from the unitarization is very large while in others it
is quite moderate. All this is calculated without any free
parameter. Therefore a comparison of the lattice results to those
from our calculation would provide a very non-trivial test of the
presence of non-perturbative hadron--hadron interactions.

\section{Chiral extrapolation}
\label{sec:chextra}

\begin{figure*}[t]
\vglue-10mm
\includegraphics[width=1.\textwidth]{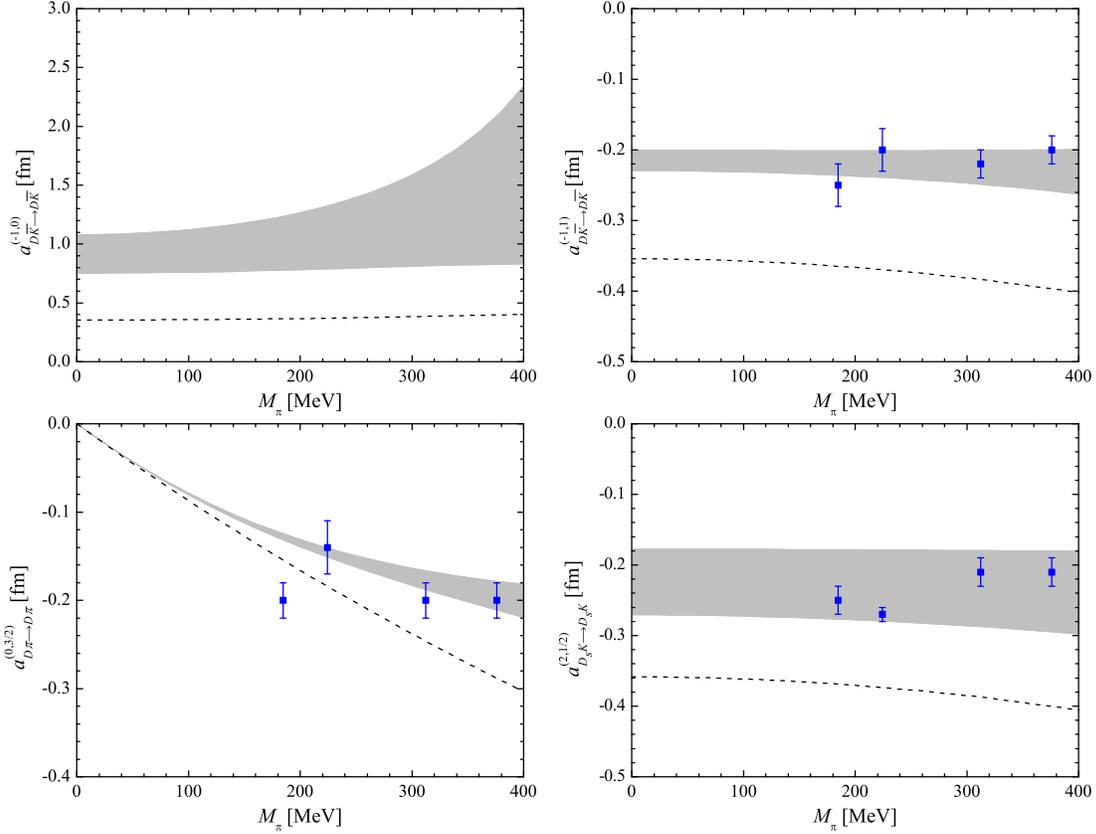}
\vglue-12mm \caption{Chiral extrapolation for the leading order
results (dashed lines) and the full UChPT calculation (bands)
compared with the lattice data.\label{fig:chex}}
\end{figure*}

The lattice results used here are calculated with finite lattice
spacing and at unphysical masses of the $u$ and $d$ quarks --- the
masses of the strange and charm quarks are chosen at their physical
values. In the lattice calculation of the charmed meson--light
hadron interactions~\cite{Liu:2008rz}, the lattice spacing is
$b=0.12$~fm. The mass of the $s$ quark is 80~MeV, which is
consistent with its physical mass, and four values are chosen for
the masses of the $u$ and $d$ quarks which are 11~MeV, 16~MeV,
32~MeV and 48~MeV, respectively. Evidently, these are larger than
the physical $u \, (d)$ quark mass. To compare the lattice results
with the  physical quantities, one needs to do extrapolations. Here
we only discuss the chiral extrapolation of the results from the
unphysical quark masses to the physical quark mass. In
Ref.~\cite{Liu:2008rz}, the authors parameterize the scattering
lengths using the first non-vanishing term in the chiral expansion.
They used $c_1+c_2M_\pi/F_\pi$ to extrapolate the $I=3/2$ $D\pi$
scattering length, and $c_1+c_2M_\pi^2/F_\pi^2$ to extrapolate the
$D_s\pi(K)$ and the $I=1$ $D{\bar K}$ \footnote{In
Ref.~\cite{Liu:2008rz} this channel was
  called $DK$.  However, the channel calculated is $[{\bar
    c}u][{\bar s}u]$, i.e. $\bar D^0K^+$~\cite{email}.} scattering
lengths. The values they got are shown in the last column of
Table~\ref{tab:a0} and Fig.~\ref{fig:slphy}.

The lattice results for the scattering lengths at unphysical quark
masses are given at several chosen values of $M_\pi/F_\pi$, which
were taken from Ref.~\cite{WalkerLoud:2008bp}. Therefore, we take
the corresponding values of unphysical $M_\pi$ from the same paper
for the lattice data points shown in Fig.~\ref{fig:chex}. Although
the masses of the charmed mesons have been expanded to one loop
order~\cite{Jenkins:1992hx}, for our purpose we only need to work to
${\mathcal O}(M_\pi^2)$. From the Lagrangian given in
Eq.~(\ref{eq:L2str}), modulo an overall unmeasurable term
proportional to $\langle\chi_+\rangle$ which can be absorbed into
the large--$N_c$ suppressed $h_0$ term, one gets for the NLO
correction to the mass squares of the  charmed mesons
\ba%
\label{eq:md}
\delta M_D^2 = 4h_1B\hat{m} ~, \
\delta M_{D_s}^2 =
4h_1B {m_s},
\ea%
where $\hat{m}=(m_u+m_d)/2$. Since the lattice
calculations~\cite{Liu:2008rz} use a value for $m_s$ which is close to
the physical $s$ quark mass, we will use the physical mass of the
$D_s$. From the first identity in Eq.~(\ref{eq:md}), the mass of the
$D$ meson up to NLO can be written as
\be%
M_D = \overset{_\circ}{M}_D + h_1 \frac{M_\pi^2}{\overset{_\circ}{M}_D}
\ee%
using $M_\pi^2=2B\hat{m}$. Using $M_K^2=B(\hat{m}+m_s)$, one gets a
similar expression for the kaon mass up to ${\mathcal O}(M_\pi^2)$ as
\be%
M_K = \overset{_\circ}{M}_K + \frac{M_\pi^2}{4\overset{_\circ}{M}_K},
\ee%
where $\overset{_\circ}{M}_K$ denotes the mass of the kaon in the
SU(2) chiral limit $m_u=m_d=0$, i.e. $ \overset{_\circ}{M}_K^{_{_{\
2}}} = B m_s$. With the above expansion of $M_D$ and $M_K$, one is
ready to extrapolate the results given in Section~\ref{sec:slphy} in
the physical world to the world with unphysical quark masses.
Assuming the subtraction constant to be $M_\pi$--independent, which
is true to NLO, the chiral extrapolations of the results using UChPT
for the same channels as in Fig.~\ref{fig:slphy} are shown in
Fig.~\ref{fig:chex} together with the lattice data. The description
of the lattice data is rather good. The width of the band reflects
the remaining freedom in the choice of the parameters which is taken
as an estimate of higher order effects. For comparison with the
dashed line we also show the result to leading order. Although
influencing the magnitude of the scattering lengths in all channels,
only in the $(0,3/2)$ $D\pi$ channel the higher order effects also
changed the pion mass dependence,
since only in this channel the changes in the light quark masses are
not dwarfed by the large strange quark mass.

\begin{figure*}[t]
\hspace{7mm}\includegraphics[width=0.5\textwidth]{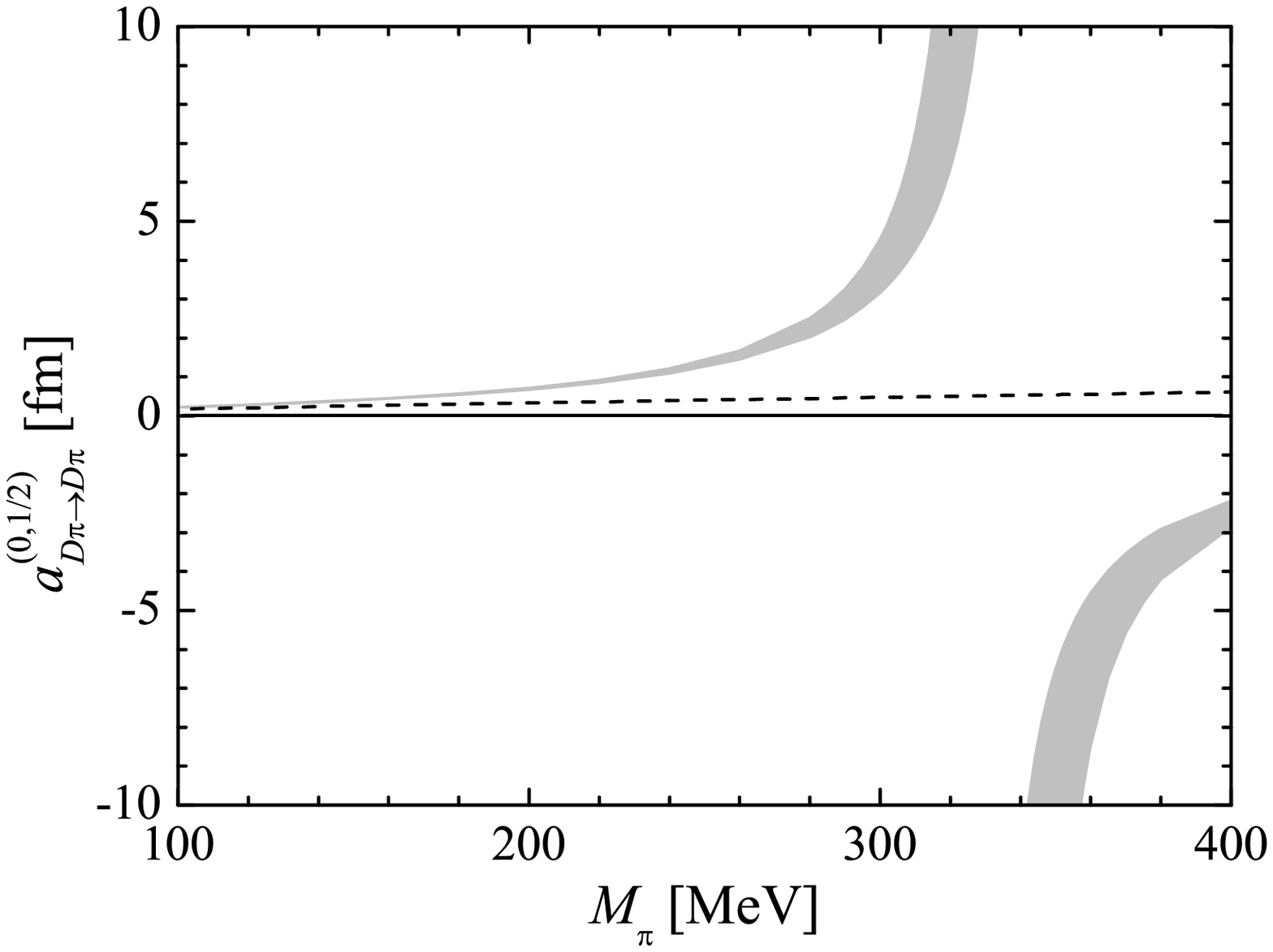}\hspace{-14mm}
\includegraphics[width=0.5\textwidth]{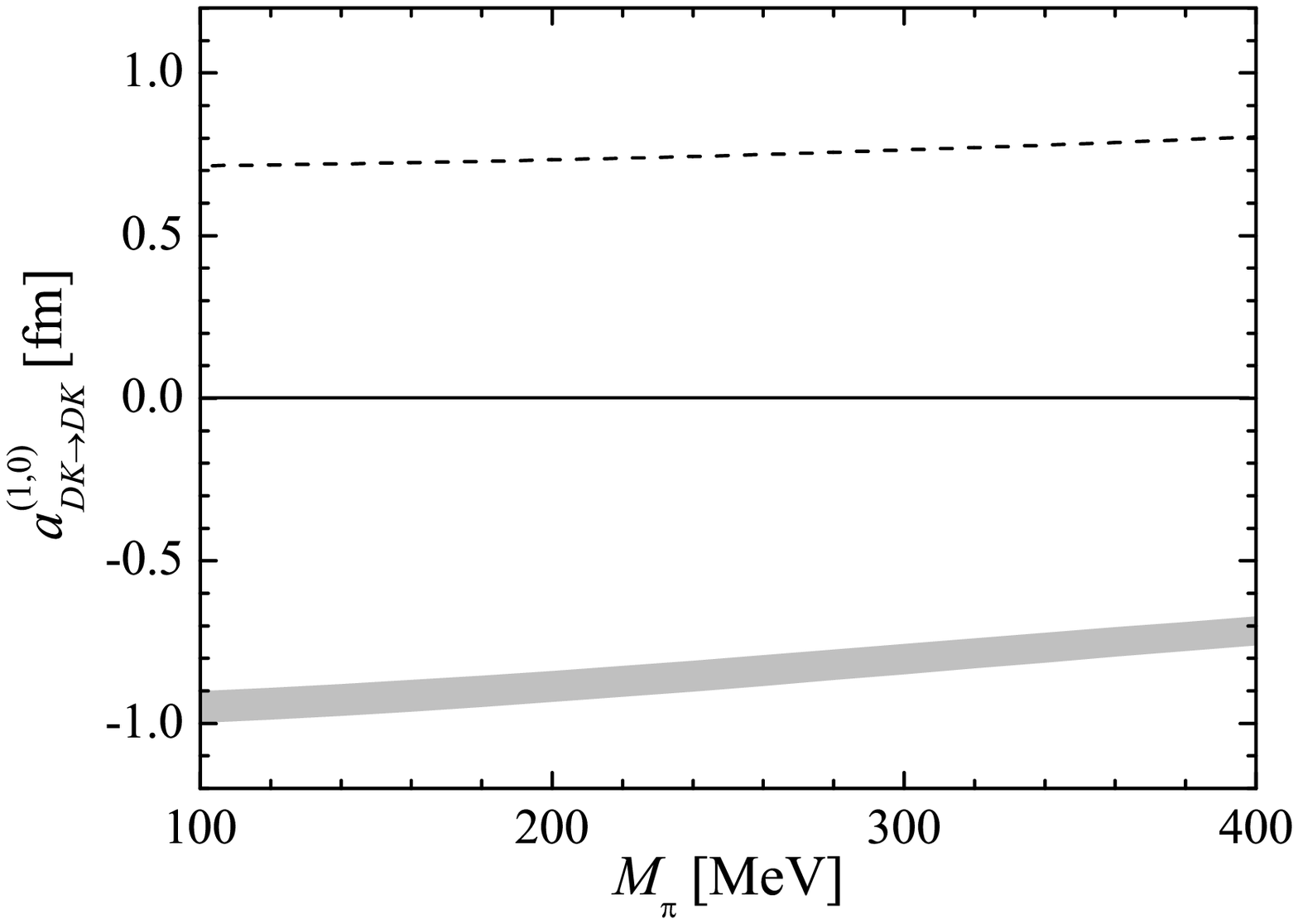}
\vglue-0mm \caption{Chiral extrapolation for the leading order
results (dashed lines) and the full CUChPT calculation (bands) for
the $(0,1/2)$ $D\pi$ and the $(1,0)$ $DK$
channels.\label{fig:chexcc}}
\end{figure*}
As discussed before, in the $(0,1/2)$ $D\pi$ and the $(1,0)$ $DK$
channels,  poles get generated through the unitarization
 which govern
the low-energy physics. Their presence  is made apparent in
Fig.~\ref{fig:chexcc}, which gives the quark mass dependence of the
scattering lengths for these channels predicted in CUChPT. To do the
calculations
 the mass of the $\eta$, which enters the coupled-channel
calculations, was expanded as
\be%
M_\eta = \overset{_\circ}{M}_\eta +
\frac{M_\pi^2}{6\overset{_\circ}{M}_\eta} \  .
\ee%
The strength of the attractive interaction in the $(0,1/2)$ $D\pi$
channel is measured by $M_\pi$. When $M_\pi$ is increased to some
value, the interaction becomes strong enough to form a bound state.
The structure shown in the left panel of Fig.~\ref{fig:chexcc} is a
nice illustration of the physics discussed in
Subsection~\ref{sec:res}. In the $(1,0)$ $DK$ channel, on the other
hand, the strength of the interaction is controlled by $M_K$, which
is changed only moderately when the $u,d$ quark masses get
increased. As a result also the scattering length shows a weak
$M_\pi$ dependence.

\section{Summary}
\label{sec:sum}

In this paper we calculated the scattering lengths for all the
Goldstone boson--$D$-meson scattering channels using both chiral
perturbation theory as well as a unitarized version of it. We found
that for some channels the unitarization produced new singularities
in the S--matrix, in some cases as virtual states or resonances, in
some cases as bound states. One of these dynamically generated
singularities can be identified with the experimentally well
established $D_{s0}^*(2317)$, that was identified as a hadronic
molecule in various
works~\cite{eef,Kolomeitsev:2003ac,Guo:2006fu,oset}. However, we
found more non-perturbative effects and demonstrated that a
determination of scattering lengths on the lattice provides an ideal
tool to investigate them.


\subsection*{Acknowledgments}

We would like to thank Liuming Liu and Kostas Orginos for useful
discussions. This work is partially supported by the Helm\-holtz
Association through funds provided to the virtual institute ``Spin
and strong QCD'' (VH-VI-231) and by the DFG (SFB/TR 16, ``Subnuclear
Structure of Matter''). We also acknowledge the support of the
European Community-Research Infrastructure Integrating Activity
``Study of Strongly Interacting Matter'' (acronym HadronPhysics2,
Grant Agreement n. 227431) under the Seventh Framework Programme of
EU.


\begin{appendix}
\section{Isospin relations}
\label{app:isore}
\renewcommand{\theequation}{\thesection.\arabic{equation}}
\setcounter{equation}{0}

In the present work, we do not consider isospin violation. In this
Appendix, we give the relations between the scattering amplitudes in
isospin basis and in particle basis which are useful in derivations.
Before that, let us give the definition of Mandelstam variables and
the phase convention for the isospin eigenstates \be
s=(p_1+p_2)^2,\quad t=(p_1-p_3)^2,\quad u=(p_1-p_4)^2~, \ee and
\begin{eqnarray}\nonumber
\left|\pi^+\rang&=&-\left|1,+1\rang,~ \left|{\bar K}^0\rang=-\left|\frac12,+\frac12\rang, \\
\left|D^+\rang&=&-\left|\frac12,+\frac12\rang.
\end{eqnarray}
The two numbers in $\left|\cdots \rang$ on the r.h.s. are $I,I_3$.
All the other states are defined with
 a positive sign. Then employing isospin symmetry and crossing
symmetry, one can get the following isospin relations:



\ba%
V^{(-1,1)}_{D\bar K\to D\bar K}(s,t,u) \! &= & \! V_{D^0K^-\to D^0K^-}(s,t,u), \\
V^{(-1,0)}_{D\bar K\to D\bar K}(s,t,u) \! &= & \! 2V_{D^+K^+\to D^+K^+}(u,t,s)
\nonumber \\ \! &- & \! V_{D^0K^-\to D^0K^-}(s,t,u), \\
V^{(0,3/2)}_{D\pi\to D\pi}(s,t,u) \! &= & \! V_{D^+\pi^+\to D^+\pi^+}(s,t,u), \\
V^{(0,1/2)}_{D\pi\to D\pi}(s,t,u) \! &= & \! {\frac32}V_{D^+\pi^+\to D^+\pi^+}(u,t,s)
\nonumber \\
&- & \! {\frac12}V_{D^+\pi^+\to D^+\pi^+}(s,t,u),\\
V^{(0,1/2)}_{D\eta\to D\eta}(s,t,u) \! &= & \! V_{D^+\eta\to D^+\eta}(s,t,u), \\
V^{(0,1/2)}_{D_s\bar K\to D_s\bar K}(s,t,u) \! &= & \! V_{D_s^+K^+\to D_s^+K^+}(u,t,s), \\
V^{(0,1/2)}_{D\eta\to D\pi}(s,t,u) \! &= & \! \sqrt{3}V_{D^0\eta\to D^0\pi^0}(s,t,u), \\
V^{(0,1/2)}_{D_s\bar K\to D\pi}(s,t,u) \! &= & \! \sqrt{3}V_{D_s^+K^-\to D^0\pi^0}(s,t,u), \\
V^{(0,1/2)}_{D_s\bar K\to D\eta}(s,t,u) \! &= & \! V_{D_s^+K^-\to D^0\eta}(s,t,u),  \\
V^{(1,0)}_{DK\to DK}(s,t,u) \! &= & \! 2V_{D^0K^-\to D^0K^-}(u,t,s)
\nonumber \\
&- & \! V_{D^+K^+\to D^+K^+}(s,t,u), \\
V^{(1,0)}_{D_s\eta\to D_s\eta}(s,t,u) \! &= & \! V_{D_s^+\eta\to D_s^+\eta}(s,t,u), \\
V^{(1,0)}_{D_s\eta\to DK}(s,t,u) \! &= & \! -\sqrt{2}V_{D_s^+\eta\to D^0K^+}(s,t,u), \\
V^{(1,1)}_{D_s\pi\to D_s\pi}(s,t,u) \! &= & \! V_{D_s^+\pi^0\to D_s^+\pi^0}(s,t,u), \\
V^{(1,1)}_{DK\to DK}(s,t,u) \! &= & \! V_{D^+K^+\to D^+K^+}(s,t,u), \\
V^{(1,1)}_{DK\to D_s\pi}(s,t,u) \! &= & \! \sqrt{2}V_{D_s^+K^-\to D^0\pi^0}(u,t,s), \\
V^{(2,1/2)}_{D_sK\to D_sK}(s,t,u) \! &= & \! V_{D_s^+K^+\to D_s^+K^+}(s,t,u), %
\ea%
where the superscripts mean $(S,I)$ with $S (I)$ representing the
total strangeness (isospin) of the two--meson system.

\end{appendix}


\end{document}